

\magnification=1200\baselineskip=0.20truein

\hsize=6.3truein

\centerline{ CANCELLATION OF INFRARED DIVERGENCES IN THERMAL QED }
\bigskip
\centerline{ H. Arthur Weldon$^{*}$}
\bigskip
\centerline{ Department of Physics}

\centerline{ West Virginia University}

\centerline{ Morgantown, WV, 26506-6315}

\vskip1truein
\noindent As a preliminary step, the radiation produced by a classical
charged current coupled to a quantized $A_{\mu}$ is solved. To each order
 in $\alpha$, all infrared
divergences cancel between the virtual $\gamma$'s and the real $\gamma$'s
 absorbed from the
plasma or emitted into the plasma.  When all
orders of perturbation theory are summed, the finite answer predicts a
suppression
 of radiation with
$\omega< \alpha T$. The analysis of QED then consists of two steps. First, a
general
 probability at
$T\neq 0$ is organized so that all the virtual $e^{\pm},\gamma$ are in the
amplitudes
  and all the
real
  $e^{\pm},\gamma$ are in the phase space
integrations. Next, the cancellations of IR divergences
between virtual and real are demonstrated.

 \vskip0.5truein

\noindent (Talk presented at Quark Matter '93)

\vskip3.5truein
\medskip
\noindent \hrule width2truein
\smallskip
\noindent  $^{*}$Supported in part by the U.S. National Science Foundation
under grant
PHY-9213734.

\vfill\eject

\noindent {\bf 1. INTRODUCTION}

In studies ofthe quark-gluon plasma to be produced in ultrarelativistic heavy
 ion colisions and,
more generally, in studies of field theory at finite temperature, a central
 concern is how the
cancellation of infrared divergences comes about. Infrared divergences occur
 when
an on-shell particle ($p^{2}=m^{2}$) emits a massless particle ($k^{2}=0$).
The resulting propagator
is $[m^{2}-(p-k)^{2}]^{-1}=[2p\cdot
k]^{-1}=[2(E-|\vec{p}|\cos\theta)|\vec{k}|]^{-1}$ and so
amplitudes for one-photon emission behave like $\sim 1/k$ at low energy.
At $T=0$ this behavior leads to  logarithmic divergences, $\int dk/k$, for
each real and virtual
photon. At $T\neq 0$ it leads to linear divergences,   $\int dk n_{B}/k$,
because of the Bose-Einstein
function  $n_{B}=1/[\exp(k/T)-1]$. This paper discusses how all the infrared

 divergences cancel
even when $T\neq 0$.

\medskip
\noindent {\bf 2. IR CANCELLATION FOR A CLASSICAL CURRENT}

\def\leftdisplay#1$${\leftline{\noindent$\displaystyle{#1}$}$$}
	\everydisplay{\leftdisplay}

	\everydisplay{\leftdisplay}

\noindent {\bf 2.1  Bremsstrahlung to Order $\alpha$}

If a charged particle scatters while passing through a fixed-temperature
plasma,
it will radiate.
If the radiated energy $\omega$ is much smaller than the energy transfer in the
collision,
then the inelastic cross section factors:
$$ 2\omega{d\sigma\over d^{3}k\ dq^{2}}\approx 2\omega {dP(q^{2})
\over d^{3}k}\ {d\sigma\over dq^{2}}
\hskip4.0truein (1)$$
so that the collison cross section $d\sigma/dq^{2}$ is independent of the
photon
energy-momentum ($\omega=|\vec{k}|$).
To first order in $\alpha$ the probability of radiating is
$$2\omega{dP_{1}\over d^{3}k}=\sum_{\rm pol}|\epsilon_{\mu}\cdot J^{\mu}|^{2}
\ [1+n_{B}(\omega)]/(2\pi)^{3}\hskip3.1truein (2)$$
When the radiated energy $\omega$ is small,  the current has a universal form
regardless of the spin of the charged particle:
$$J^{\mu}(k)=ie\left({p^{\prime\mu}\over p^{\prime}\cdot k}-
{p^{\mu}\over p\cdot k}\right)
\ e^{-k/2\Lambda}\hskip3.55truein (3)$$
 Here $\Lambda$ is a momentum cutoff that
is necessary later.  The radiation is mostly parallel to $\vec{p}$ or
 $\vec{p}^{\ \prime}$.
When integrated  over angles the result is
$${dP_{1}\over d\omega}={A\over\omega} [1+n_{B}(\omega)]\
 e^{-\omega/2\Lambda},\hskip1truein
A(p\cdot p^{\prime})
= {\alpha\over \pi}\left[{1\over v}\ln({1+v\over 1-v})-2\right]
\hskip0.6truein (4)$$
where $v$  is defined by $p\cdot p^{\prime}=m^{2}(1-v^{2})^{-1/2}$.
 At large momentum transfer ($Q\gg
m$), the behavior is $\pi A\approx 4\alpha \ln(Q/m)$. Except for the
statistical factor $n_{B}$, (4)
 is a classical formula. When $\omega\ll T$ this predicts
$dP_{1}/ d\omega\approx AT/\omega^{2}.$
This is totally unphysical because the   energy radiated in low energy
modes, below some $E_{\rm
max}$, would be infinte:  $\int_{0}^{E_{\rm max}} d\omega \omega
 dP_{1}/d\omega=\infty.$

\vfill\eject
\noindent {\bf 2.2 Bremsstrahlung to All Orders in $\alpha$}

To improve upon (4),  couple the classical current (3) to the quantized
 radiation field $A_{\mu}$.
The generating  function for all  multi-photon amplitudes can be obtained by
 a Gaussian functional
integration. From this one obtains the exact multi-photon amplitudes ${\cal
M}$.
The probability that any number $n$ of real photons in the plasma will radiate
a
 \underbar{net}
energy $\omega$  is $${dP\over d\omega}=\sum_{n=1}^{\infty}
\int d\Phi_{1}...d\Phi_{n}
\delta(k_{1}^{0}+...k_{n}^{0}-\omega)
{1\over n!}\sum_{\rm pol}|{\cal M}(k_{1},...k_{n})|^{2}\hskip1.7truein (5)$$
$$d\Phi_{i}\equiv{d^{4}k_{i}\over
(2\pi)^{3}}\delta(k_{i}^{2})\ [\theta(k_{i}^{0})+n_{B}(|\vec{k}_{i}|)]
\hskip3.5truein (6)$$
This weights photon emission ($k^{0}>0$)  with the statistical factor
$1+n_{B}$; and
photon absorption ($k^{0}<0$)  with the factor $n_{B}$.

 Each amplitude ${\cal M}$ is infrared divergent from closed loops of virtual
photons.
  Each  integration $d\Phi$ over the real photons is infrared divergent.
The virtual and real contributions exponentiate to give
$${dP\over d\omega}=\int_{-\infty}^{\infty}{dz\over 2\pi}e^{-i\omega z}
\exp[\overline{R}(z)]\hskip 3.9truein (7)$$
$$\overline{R}(z)
=\int{d^{3}k\over 2k(2\pi)^{3}}J_{\mu}(k)J^{\mu}(k)
\left([1+n_{B}]e^{ikz}
+n_{B}e^{-ikz}-[1+2n_{B}]\right)\hskip1.15truein (8)$$
The term proportional to $1+n_{B}$ represents stimulated emission; the
term proportional to $n_{B}$
represents  absorption; the term proportional to $1+2n_{B}$ represents
the virtual
photons (emitted and absorbed). At small $k$, $J^{\mu}\sim 1/k$ and
$n_{B}\sim 1/k$. Nevertheless,
all infrared divergences cancel in (8)  and give a finite result.
It is possible to  compute
$\overline{R}(z)$ and to compute the Fourier transform (7). The
final result is
$${dP\over d\omega}=|\Gamma({A\over 2}+i{\omega\over 2\pi T})|^{2}\
{e^{\omega/2T}e^{-|\omega|/\Lambda}\over 4\pi^{2}T\ \Gamma(A)}\
 \left({2\pi T\over
\Lambda}\right)^{A}\hskip2.6truein (9)$$
where $A$ is the same function (4) as before. The most interesting feature
 of (9) is the appearance
of two dimensionless scales:  $A\ll 1$ and $\omega/T$. If
$A\pi T\ll\omega$ then
$${dP\over d\omega}\approx{A\over\omega} [1+n_{B}(\omega)]\
 e^{-\omega/\Lambda}
\hskip0.3truein (A\pi T\ll\omega)\hskip2.85truein (10)$$
which agrees with the first-order result (4).
However this does not apply  at $\omega\ll T$. At small energy,
$${dP\over d\omega}\approx{AT\over \omega^{2}+(A\pi T)^{2}}\hskip0.9truein
(\omega\ll T)\hskip3.0truein (11)$$
Naturally (10) and (11) agree  in the region of overlap. Surprisingly,
 $dP/d\omega$ is constant
for $\omega\ll A\pi T$ rather than increasing like $1/\omega^{2}$.
The  interpretation of the small $\omega$ suppression is that the
quantity $2 A\pi T=\Gamma_{r}$ is a
damping rate produced by the radiation reaction as required by unitarity [1].

\medskip
\noindent {\bf 3.  REAL \& VIRTUAL PARTICLES IN THERMAL FIELD THEORY}

We now set aside the semiclassical approximation and turn to the
full quantum field theory.
Each species of particle has four types of propagator in thermal
field theory, e.g.
$S_{ab}$ for electrons and $D_{ab}^{\mu\nu}$ for photons
with $a,b=1\ {\rm or}\ 2$. It is possible to
rewrite thermal probabilities as squares of amplitudes that contain
only $S_{11}$ and
$D_{11}^{\mu\nu}$, integrated over physical phase space.

For definiteness, consider the process $e^{-}(p_{1})$ + plasma $\to$  anything,
with rate
$$R(p_{1})=\sum_{F,I}|<F|C|I>|^{2}
\ {e^{-\beta E_{I}}\over Z }\hskip1truein  C\equiv
[S,b^{\dagger}(\vec{p}_{1})]\hskip1.4truein (12)$$
Using completeness and thermofield dynamics [2] one can write this as
$$R(p_{1})=<0(\beta)|C^{\dagger}C|0(\beta)>=\sum_{F}
|<F(\beta)|C|0(\beta)>|^{2}\hskip 2.0truein (13)$$
where $|F(\beta)>$ is a compete set of thermal modes. These correspond
 to real particles:
$$R(p_{1})=\sum_{\ell=2}^{\infty} \int d\Psi_{2}...d\Psi_{\ell}
\sum_{n=0}^{\infty}\int d\Phi_{1}...d\Phi_{n}
{1\over n!}|M_{\ell,n}(p_{1},...p_{\ell};k_{1},...k_{n})|^{2}
\hskip0.95truein (14)$$
where ${\cal M}_{\ell n}$ is the amplitude for  $\ell$ real $e$'s or
 $\overline{e}$'s (including the
initial one) and $n$ real $\gamma$'s. By charge conservation, $\ell$
is even. Here $d\Psi$ is the
fermion phase space similar to the photon phase space $d\Phi$ in (6).
The explicit amplitude has the form
$$M_{\ell,n}=\sum_{I}<I|[a^{\#},[a^{\#},...\{b(\vec{p}_{2}),
[S,b^{\dagger}(\vec{p}_{1})]\}]]|I>
\ {e^{-\beta E_{I}}\over Z}\hskip1.75truein (15)$$
These amplitudes are related to Green functions  that do not have
the usual $\exp(-\beta H)/Z$
but rather $S\exp(-\beta H)/Z$.
 Consequently $ M_{\ell,n}$ has \underbar{no}  ``cut" propagators
$D_{21}= D_{12}$ or $S_{21}=-S_{12}$. It is constructed entirely of
 $D_{11}(k)$ and $S_{11}(p)$. See also [3,4].

\medskip

\noindent {\bf 4. 	CANCELLATION OF INFRARED DIVERGENCES }

To analyze the rate (14) one can repeat the  analysis of Yennie,
Frautschi, and  Suura [5] with
slight modifications.
Virtual photons inside the amplitude ${\cal M}_{\ell n}$
can cause an IR divergence when
they are on-shell and attached to ``external" fermions
 $p_{1},...p_{\ell}$.
Each end of a photon line attached to an ``external"
fermion gives a multiplicative factor
$ e2p^{\mu}/ 2p\cdot k$ plus non-IR terms.
 At each order of $\alpha$, the amplitude has a maximum IR
divergence $\alpha^{p}$ plus non-leading divergences
 $\alpha^{p-1}$, $\alpha^{p-2}$,....
 When summed to all orders, the leading and
 non-leading divergences
 all exponentiate:
$$M_{\ell,n}=\exp(V)\ M_{\ell,n}^{\rm fin}\hskip0.5truein
V=\int {d^{3}k\over (2\pi)^{3}2k}[1+2n_{B}]{1\over 2}
\sum_{a,b=1}^{\ell}{e_{a}e_{b}\ p_{a}\cdot p_{b}\over
(p_{a}\cdot k)(p_{b}\cdot k)}\
e^{-k/\Lambda}\hskip0.1true in(16)$$
$V$ by itself is IR divergent.
For real photons, infrared divergences   occur in the integrations
$d\Phi$ in (14). For fixed fermion
momenta, define
  $$R_{\ell}(p_{1},...p_{\ell})=\sum_{n=0}^{\infty}\int
d\Phi_{1}...d\Phi_{n}{1\over
n!}|M_{\ell,n}|^{2}\hskip3.0truein(17)$$ Since $d\Phi\sim
kdk[\theta(k^{0})+ n_{B}]$, divergences
arise when ${\cal M}_{\ell n}\sim 1/k$.  When $n$ real photon lines
are attached to the ~external"
fermions in all possible ways, they each give multiplicative factors
 $ e2p^{\mu}/2p\cdot k$
plus non-IR terms.
The contribution to (17) of $n$ real photons
has an IR divergence $\alpha^{n}$ plus non-leading divergences
$\alpha^{n-1}$, $\alpha^{n-2}$, .. 1.
When summed  over $n$, the leading and non-leading
 divergences all exponentiate to give:
 $$R_{\ell}(p_{1},...p_{\ell})=\sum_{m=0}^{\infty}\int_{-\infty}^{\infty}
 d\omega{dP\over d\omega}
\int d\Phi_{1}...d\Phi_{m}
\beta_{\ell,m}(k_{1},...k_{m})/m!\hskip1.5truein (18)$$
All IR divergences are contained in the quantity
$${dP\over d\omega}=\int_{-\infty}^{\infty}{dz\over 2\pi}e^{-i\omega z}
\exp[\overline{R}(z)]\hskip 3.8truein (19)$$
$$\overline{R}(z)
=\int{d^{3}k\over 2k(2\pi)^{3}}\sum_{a,b=1}^{\ell}{e_{a}e_{b}p_{a}
\cdot p_{b}\over (p_{a}\cdot k)
(p_{b}\cdot k)}
\left([1+n_{B}]e^{ikz}
+n_{B}e^{-ikz}-[1+2n_{B}]\right)e^{-k/\Lambda} (20)$$
This integration is IR finite because of the delicate cancellation between the
virtual photons and the
real photons (emitted +
absorbed). The calculation is the same as in Sec. 2 with the result
$${dP\over d\omega}=|\Gamma({A\over 2}+i{\omega\over 2\pi T})|^{2}\
{e^{\omega/2T}e^{-|\omega|/\Lambda}\over 4\pi^{2}T\Gamma(f)}({2\pi T\over
\Lambda})^{A}\hskip2.8truein (21) $$
but now $A$ depends on all the ``external" fermion momenta:
$$A\equiv
-\sum_{a,b=1}^{\ell}{e_{a}e_{b}\over 8\pi^{2}}{1\over v_{ab}}
 \ln\left({1+v_{ab}\over
1-v_{ab}}\right)\quad \ge 0\hskip3.1truein (22)$$ and
$v_{ab}$, defined by $p_{a}\cdot
p_{b}/m^{2}=(1-v_{ab}^{2})^{-1/2}$,
 is relative velocity of charge $a$ in the rest frame of  $b$.

\medskip
\noindent {\bf 5. CONCLUSIONS}

For fixed momenta of the ``external" fermions, the rate (18) is  IR finite
because (a) the function
$dP/d\omega$ is  finite and (b) the integration $\int d\omega dP/d\omega$ is
finite due to the
behavior $dP/d\omega\to$ constant as $\omega\to 0$.

The full rate requires integration  over the
 thermalized fermions:
$$R(p_{1})=\sum_{\ell=2}^{\infty} \int
d\Psi_{2}...d\Psi_{\ell}\ R_{\ell}(p_{1},...p_{\ell})\hskip3.3truein (23)$$
The fermion integrations do not affect the infrared finiteness of $R(p_{1})$.
However,
there will still be Coulomb divergences  that arise when any one of the
momentum transfers
vanishes: $(p_{a}-p_{b})^{2}\to 0$. It is known that the usual $T=0$ Coulomb
divergence
$\int \sin\theta d\theta/\theta^{4}$ is reduced to logarithmic,
$\int \sin\theta d\theta/\theta^{2}$
at $T\neq 0$ due to Braaten-Pisarski resummation [6].

For definiteness, the process
$e^{-}$ + plasma $\to$ anything was treated specifically.
 For a general process $\{ A\}$ + plasma $\to \{B\}$ + anything,
where $\{A\}$ and $\{B\}$ are any sets of $e^{\pm}, \gamma$, the same
arguments apply and show that
all infrared divergences cancel. However, the logarithmic Coulomb divergence
remains.

\medskip
\noindent{\bf REFERENCES}

\item{1.} H.A. Weldon, Suppression of Bremsstrahlung at Non-zero Temperature,
submitted
 to Phys. Rev. D.

\item{2.}  H. Matsumoto, I. Ojima, and H. Umezawa, Ann. Phys. (NY) 152 (1984)
348.

\item{3.} N. Ashida, H. Nakkagawa, A. Ni\'egawa, and H. Yokata, Phys. Rev. D45
(1992) 2006
and Ann. Phys. (NY) 215 (1992) 315.

\item{4.} A. Ni\'egawa and K. Takashiba, Nucl. Phys. B370 (1992) 335.

\item{5.} D. R. Yennie, S.C. Frautschi, and H. Suura, Ann. Phys. (NY) 13 (1961)
379.

\item{6.} R.D. Pisarski, Phys. Rev. Lett. 63 (1989) 1129; E. Braten and R.D.
Pisarski, Phys. Rev.
Lett. 64 (1990) 1338.

\end